\documentclass[
twocolumn,
aps,nofootinbib,showpacs,showkeys,preprint
tightenlines,preprintnumbers,
,superscriptaddress
] {revtex4}

\usepackage{epsf,epsfig,subfigure,graphicx,amsmath,amssymb}
\usepackage{color}
\usepackage{float}

\newcommand{\dis}[1]{\begin{equation}\begin{split}#1\end{split}\end{equation}}

\newcommand\lsim{\mathrel{\rlap{\lower4pt\hbox{\hskip1pt$\sim$}}
    \raise1pt\hbox{$<$}}}
\newcommand\gsim{\mathrel{\rlap{\lower4pt\hbox{\hskip1pt$\sim$}}
    \raise1pt\hbox{$>$}}}

\newcommand\ie{{\it i.e.}~}
\newcommand\gev{\,{\rm GeV}}

\newcommand{\Z}[1]{{\bf Z}}

\begin{document}

\title{Axino mass\footnote{Talk presented at DSU 2012, Buzios, Brasil, 10-15 June 2012.}}

\author{ Jihn E. Kim }
\affiliation{
GIST College, Gwangju Institute of Science and Technology, Gwangju 500-712, Korea , }
 \author{ Min-Seok Seo}
\affiliation{
 Department of Physics and Astronomy and Center for Theoretical
 Physics, Seoul National University, Seoul 151-747, Korea  }

\begin{abstract}
I will talk on my recent works. Axino, related to the SUSY transformation of axion, can mix with Goldstino in principle. In this short talk, I would like to explain what is the axino mass and its plausible mass range. The axino mass is known to have a hierarchical mass structure depending on accidental symmetries. With only one axino, if $G_A=0$ where $G=K+\ln|W|^2$, we obtain $m_{\tilde a}=m_{3/2}$.
For $G_A\ne 0$, the axino mass depends on the details of the K\"ahler potential. I also comment on the usefulness of a new parametrization of the CKM matrix.
\end{abstract}


 \maketitle

%%%%%%%%%%%%%%%%%%%%%%%%%%%%%%%%%%%%%%%%
\section{Introduction}\label{sect:intro}

In this talk I will concentrate on the questions, ``What is axino?, What is Goldstino?," and ``What is the axino mass?"

\begin{figure}[h]
 \includegraphics[width=0.23\textwidth]{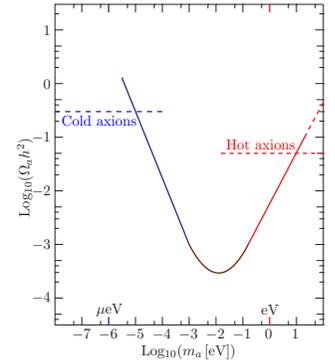}\\
\includegraphics[width=0.25 \textwidth]{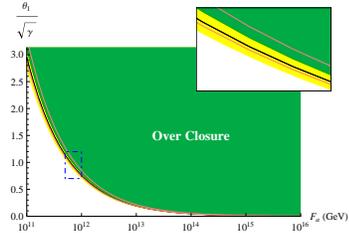}
  \caption{The axion energy density curves \cite{KimRMP10,BaeRhoa08}.}
\label{fig1:rhoa}
\end{figure}

Dark matter(DM) in the universe is the most looked-for particle(s)
in cosmology and at the LHC, and also at low temperature axion search laboratories.
The 100 GeV scale weakly interacting massive particle(WIMP) \cite{LeeWein77} and the 10-1000 $\mu$eV axion \cite{PWW83} are the most promising DM candidates.
In the top side of Fig. \ref{fig1:rhoa}, we show the cold DM(CDM) axion case \cite{KimRMP10}. For the WIMP, the idea has been originated from the the heavy neutrino case \cite{Kolb90} and pointed out in \cite{Goldberg83}.

For the case of axion, the axion potential is very flat for a large axion decay constant compared to that of a small axion decay constant, and the minimum is at the CP conserving point in the effective theory of QCD. [Note, however, that if the weak CP violation is considered, then the minimum point is shifted a bit but far below the current experimental limit on $\overline{\theta}$.]
In the evolving universe, at some temperature, say $T_1$, the classical axion field $\langle a\rangle$starts to roll down to end at the CP conserving point
sufficiently closely. This analysis constrains the axion decay
constant $f_a$ (upper bound) and the initial VEV $f_1\equiv \langle a\rangle$ of $a$ at temperature  $T_1$. The recent study \cite{BaeRhoa08} in the $\theta_1\equiv f_1/f_a$ versus $f_a$ plane is shown in the bottom side of Fig. \ref{fig1:rhoa}.
%%%%%%%%%%%%%%%%%%%%%%%%%%%%%%%%%%%%%%%%%%%%%%%%%%%%%%%%%%%%%%%%%%%%%%%%%%%%%%%%%%%
%%%%%%%%%%%%%%%%%%%%%%%%%%%%%%%%%%%%%%%%%%%%%%%%%%%%%%%%%%%%%%%%%%%%%%%%%%%%%%%%%%

Both the WIMP CDM and axion CDM contribute to the galaxy formation, and hence the naive N-body simulation cannot distinguish the WIMP or the axion formation of galaxies. In this regard, we point out the tidal torque theory in the case of axion CDM when the axions go through the Bose-Einstein condensation(BEC) before the formation era of galaxies \cite{YangSikivie10}. For the case of BEC, there can exist a net overall rotation via BEC, because  in the lowest energy state  all axions fall with the same angular momentum. On the other hand, WIMPs have an irrotational
velocity field.

Two most persuasive reasons toward the very light axion are its solution of the strong CP problem and its role in the galaxy formation. The other most conspicuous problem, the TeV scale scalar mass problem, proposes supersymmetry(SUSY) as its solution. Thus, the obvious combined solution for the strong CP problem and the scalar mass problem  needs supersymmetrization of an axion model, predicting its superpartner {\it axino}. The axino has been considered in the context of the invisible axion \cite{axinoearly}, its effects to cosmology at the eV, keV, GeV, and TeV scales \cite{axcosmologyeV,axcosmologykeV,axcosmologyGeV,axcosmologyTeV}. The axino interaction with electron, saxion effects, and the hot thermal loop contribution have been considered in \cite{axinoInt}. In all these cosmological applications, the magnitude of the axino mass is crucial \cite{AxinoRevs}.

The axino mass has been considered for the case with an accidental symmetry \cite{ChunKN92}, in plausible SUGRA models \cite{ChunLukas95}, and most recently in a general framework taking into account the axino-gravitino mixing \cite{KSeoAxinomass12}.

%%%%%%%%%%%%%%%%%%%%%%%%%%%%%%%%%%%%%%%%%%%%%%%%%%%%%%%%%%%%%%
\section{The PQ symmetry in supergravity}\label{sec:AxinoTh}

The axion in the spontaneously broken PQ is parametrized by
\begin{eqnarray}
f_a e^{ia/f_a} &=&\sum_{i} v_i e^{ia_i/f_i} \label{eq:axiondefNonSUSY}
\end{eqnarray}
which is $a  \propto \sum_{i} v_i e^{ia_i/f_i} $ in the small field approximation.
Our question is, ``With SUSY, how can we define the pseudo scalar $a$ in a spontaneously broken PQ symmetry? Or more generally, in a spontaneously broken global symmetry?" For this, it is customary to parametrise the SUSY breaking with the PQ symmetry. The prototype form is \cite{Kim84},
\begin{eqnarray}
W= Z_1 (S_1S_2-f_1^2)+Z_2 (S_1S_2-f_2^2),~~ {\rm with}~f_1\ne f_2,
\end{eqnarray}
where the chiral superfields are $Z_1, Z_2, S_1$, and $S_2$.  The axion superfield $A$ is composed of {\it axion a, saxion s}, and {\it axino} $\tilde a$,
\begin{eqnarray}
A=\frac{1}{\sqrt2}(s+ia)+\sqrt2 \tilde{a}\, \vartheta +F_A \vartheta ^2
\end{eqnarray}
where $F_A$ is auxiliary field and is not treated as an independent field. In other words, $F_A$ is expressible in terms of dynamical fields.  In SUGRA, we need the K\"ahler potential $K$ and the potential $V$  which is a function of chiral scalar fields $\phi_i$ and superfields $\Phi_i \ni \phi_i$,
\dis{
K &=\sum_{I,J}\sum_{i,j}f_I(\phi_i) g_J(\phi_j)+{\rm h.c.}+\cdots,\\
V &=\left(\sum_{A,B}\sum_{i,j}\int d^2\vartheta d^2\overline{\vartheta} p_A(\Phi_i) q_B(\overline{\Phi}_j)+{\rm h.c.}\right)+\cdots
}

\begin{figure}[t]
   \includegraphics[width=0.25\textwidth]{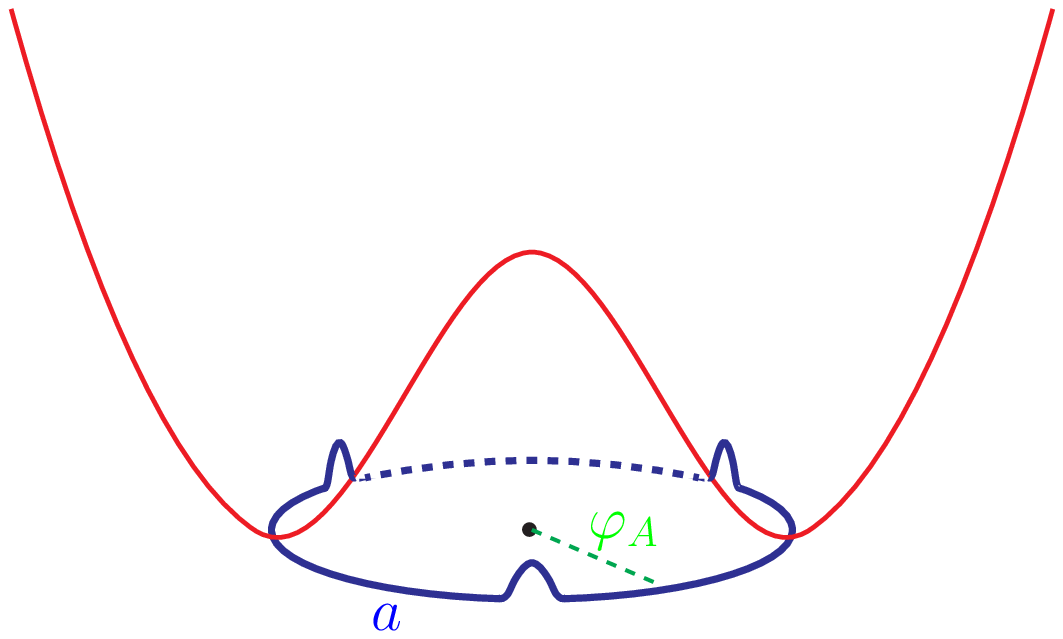}\\
\vskip 0.3cm\includegraphics[width=0.25\textwidth]{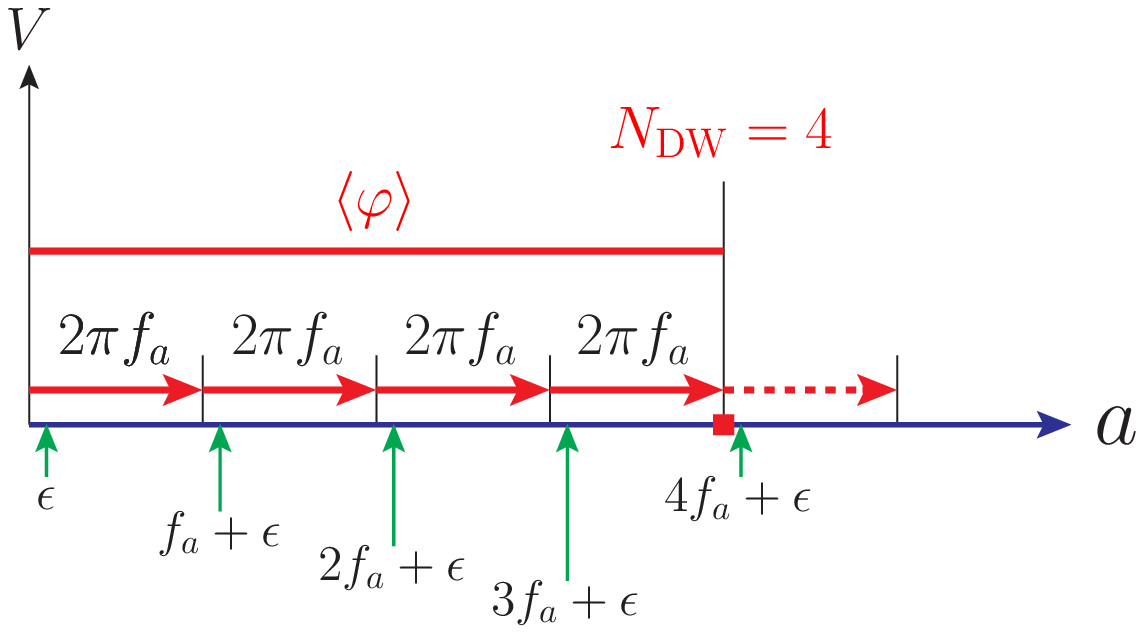}
\caption{The phase shift of the PQ charged field along the blue valley. In the upper figure, it is the valley of the Mexican hat, separated by  the axion domain wall number $N_{\rm DW}=3$. The radial field $\varphi_A$ is not the real $\vartheta^0$ part of axion superfield $A$. In the lower figure, the shift is shown for   $N_{\rm DW}=4$ \cite{AxDomainWall,KSeoAxinomass12}.} \label{fig:AxionShift}
\end{figure}

The question is how we write the field $A$ in $W$ and $K$. If the PQ symmetry is linearly realized, $A$ must come from a combination of chiral fields. In Eq. (\ref{eq:axiondefNonSUSY}), the axion field appears in the exponent in the linear realization, \ie as the phases of the PQ charge nonzero fields. In Fig. \ref{fig:AxionShift}, we show how the axion field shifts under the PQ transformation.
So, the $\vartheta^0$ component of $A$ cannot be a radial field.
%%%%%%%%%%%%%%%%%%%%%%%%%%%%%%%%%%%%%%%%%%%%%%%%%%%%%%%%%%%%%%%%%%%%%%%%%%%%%%%%%%%
Thus, there is a need to introduce the radial field corresponding to $A$. Let us call the radial fields as $\varphi$ type fields. For example, the radial field corresponding to $A$, $\varphi_A$, is composed of two real fields $\rho_{\perp}$ and Im$\,\varphi_A$. So is any radial field corresponding to the phase shift $i$: $\varphi_i$. Their VEVs are $\langle\varphi_A \rangle=V_a$ and $\langle\varphi_i\rangle=v_i$. Also, $\vartheta^0$ component of $A$ is composed of two real fields $s$ and $a$, and its VEV is defined to be vanishing  $\langle A\rangle=0$. The axion decay constant is the VEV of the radial field $\varphi_A$.

Thus, the SUSY generalization of Eq. (\ref{eq:axiondefNonSUSY}) is
\begin{eqnarray}
\Gamma_a \varphi_a e^{A/f_a}=\sum_{i}\frac{v_i}{V_a}\Gamma_i \varphi_i\,  e^{A/f_a}
\label{eq:axiondefSUSY}
\end{eqnarray}
where $\Gamma$'s are the PQ charges of the chiral fields, and $\varphi_i$ appears as the coefficient outside the exponent.

The model-independent axion in superstring models is combined with the dilaton to make a supermultiplet \cite{Svrcek06},
$D= \frac{1}{g^2}+i\frac{a_{MI}}{8\pi M_P} \longrightarrow
s +\frac{f_{MI}}{8\pi }e^{ia_{MI}/f_{MI}}$,
where $f_{MI}\sim 10^{16}$\,GeV \cite{ChoiK85}, and $\langle s\rangle\simeq 2M_P$ is not the $\varphi$ type field. Because the corresponding U(1) is gauged,  $a_{MI}$ is absorbed to the U(1) gauge boson, and the U(1) symmetry remains as a global PQ symmetry below the scale $f_{MI}$. Since this anomalous model-independent axion is given as a nonlinear form in string models, there is no accompanying $\varphi$ type field. Below $f_{MI}$, the resulting pseudo-Goldstone boson will accompany  a $\varphi$ type field. Probably, this model independent axion is the only place for the axion not accompanying its $\varphi$ type field in SUGRA axion models. Maybe, supersymmetrization of composite axion models \cite{AxComp85} encounter a similar situation.

%%%%%%%%%%%%%%%%%%%%%%%%%%%%%%%%%%%%%%%%%%%%%%%%%%%%%%%%%%%%%%%%%%%
\section{Goldstino, axion and axino}

The axion component is defined in Eq. (\ref{eq:axiondefSUSY}). So, whatever the non-vanishing PQ charge carrying F-terms are, the axion is properly defined only  by the PQ charge carrying $\vartheta^0$ terms. However, the nonvanishing F-terms define the Goldstino component.

\begin{figure}[t]
\includegraphics[width=0.2\textwidth]{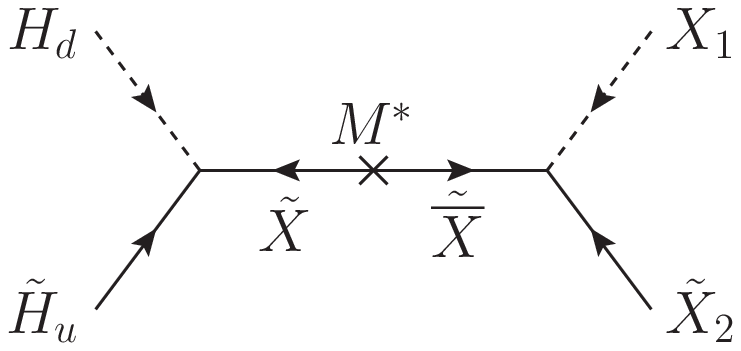}\\
  \vskip 0.5cm  \includegraphics[width=0.3\textwidth]{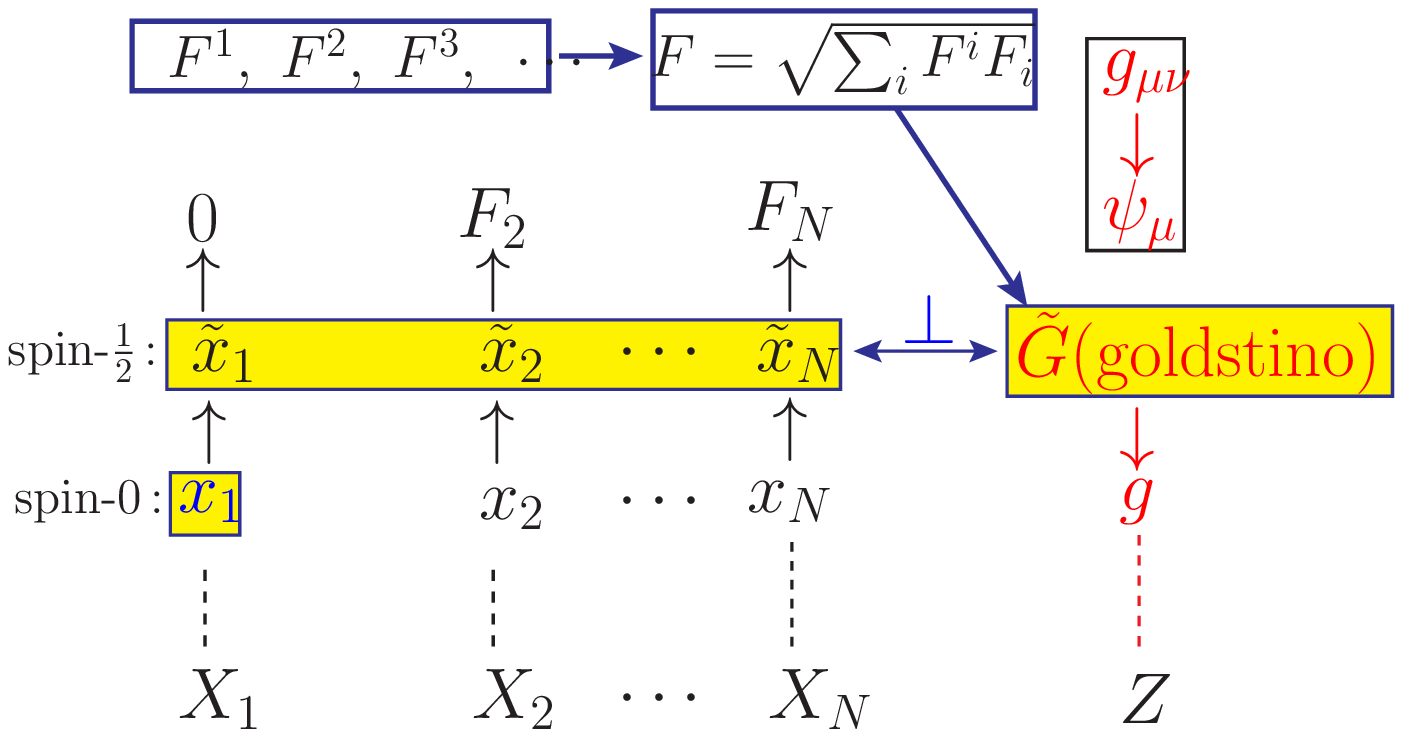}
\caption{The axino and Goldstino. In the top figure, the Kim-Nilles mechanism is shown as superpotential terms.  In the bottom figure, the origin of axion and Goldstino are shown.} \label{fig:KNandAxino}
\end{figure}

When fields carrying the PQ charges develop VEVs, the PQ symmetry is broken. When fields develop F-terms, SUSY is broken. In addition, if the F-term carries the nonvanishing PQ charge, then the PQ symmetry is also broken. However, the F-term is auxiliary, and hence the PQ symmetry breaking can only be discussed in terms of coefficient fields of $\vartheta^0$ component of $A$. This can be most succinctly presented with the following $W$ and $K$, suppressing the coupling constants,
\begin{eqnarray}
W = X_1 X_2 X +H_uH_d \overline{X}+MX\overline{X},~~~
K= \frac{H_uH_d}{M_P} X^*\,, \nonumber
\end{eqnarray}
which allow the PQ charges of the fields as, $\Gamma(H_u)=1, \Gamma(H_d)=1, \Gamma(X_1)=-1,\Gamma(X_2)=-1, \Gamma(X)=2$, and $\Gamma(\overline{X})=-2$. The Giudice-Masiero mechanism \cite{GiuMas88} uses $F^*$-term of $X^*$ in $K$ with $\mu_{\rm GM}=\frac{1}{M_P} F_X^* =\frac{1}{M_P} \frac{\partial W}{\partial X} =\frac{X_1X_2}{M_P} $. On the other hand, the Kim-Nilles employs the PQ invariant nonrenormalizable term $W= \frac{X_1X_2}{M_P}H_u H_d $, leading to $\mu_{\rm GM}=\frac{X_1X_2}{M_P}$ \cite{KimNilles}.  In the top figure of Fig. \ref{fig:KNandAxino}, the relevant Feynman diagram is shown. Thus, in the full theory, they must give the same or similar results. For the effective electroweak(EW) scale interaction, we need not consider the F-term for the global symmetry breaking.
%%%%%%%%%%%%%%%%%%%%%%%%%%%%%%%%%%%%%%%%%%%%%%%%%%%%%%%%%%%%%%%%%%%%%%%%%%%%%%%%%%%
%%%%%%%%%%%%%%%%%%%%%%%%%%%%%%%%%%%%%%%%%%%%%%%%%%%%%%%%%%%%%%%%%%%%%%%%%%%%%%%%%%
Recently, supersymmetric axion models got a lot of interest in view of the recent LHC data \cite{KimNillSeo12}.

\begin{figure}[t]
  \includegraphics[width=0.3\textwidth]{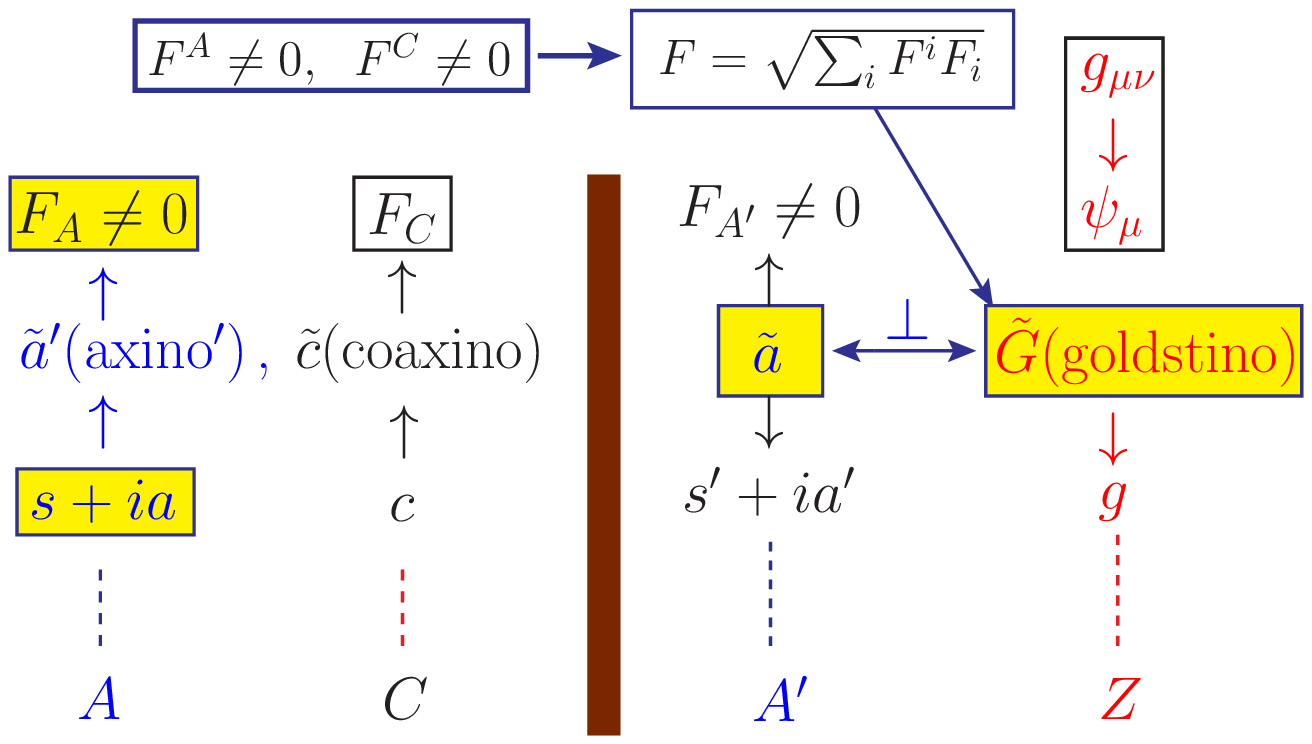}\\
 \vskip 0.5cm
    \includegraphics[width=0.2\textwidth]{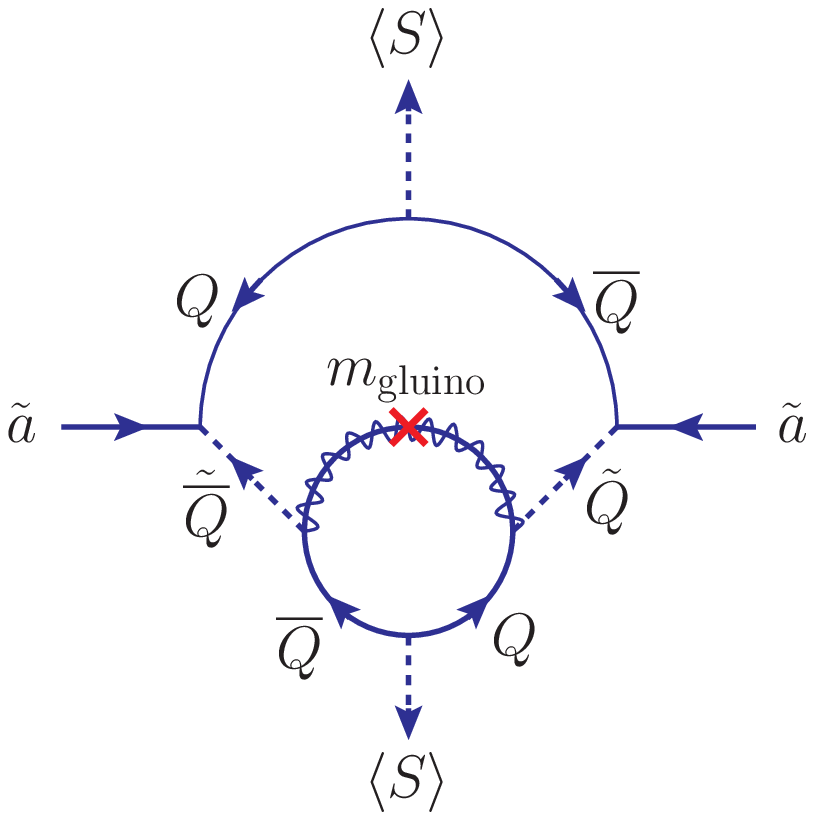}
\caption{The $F^A\ne 0$ (gravity mediation) and gaugino (gauge mediation) contributions to the axino mass.} \label{fig:FAandGaugino}
\end{figure}

Supersymmetry is spontaneously broken when the potential has nonzero VEV, $\langle V \rangle =\sum_i F^i F_i>0$ where $F^i \equiv K^{i{\bar j}}F_{\bar j}$.  Then, there should be a massless fermion, Goldstino. In supergravity, it is absorbed to the longitudinal component of gravitino $\psi_\mu$ through the super-Higgs mechanism. The Goldstino superfield, to which Goldstino belongs, can be defined by $
 Z=\sum_i \frac{F^i}{F} X_i$,
where $F=\sqrt{\sum_i F^iF_i}$which becomes the F-term of $Z$. Among $X_i$, the axion superfield is defined by the PQ charges of $X_i$. All the other chiral fields orthogonal to $A$ are called {\it coaxino} directions as shown in the bottom figure of Fig. \ref{fig:KNandAxino}. Then, we can consider two cases in which the axion superfield $A$ allows: (1)
  $F_{A} \ne 0$, or  (2) $F_{A} =0$, but $F^A\ne 0$ from K\"ahler mixing with other SUSY breaking fields.
In any case, $F^A\ne 0$ which is shown in the top figure of Fig. \ref{fig:FAandGaugino}.
%%%%%%%%%%%%%%%%%%%%%%%%%%%%%%%%%%%%%%%%%%%%%%%%%%%%%%%%%%%%%%%%%%%%%%%%%%%%%%%%%%%
%%%%%%%%%%%%%%%%%%%%%%%%%%%%%%%%%%%%%%%%%%%%%%%%%%%%%%%%%%%%%%%%%%%%%%%%%%%%%%%%%%
Definition of axion $a$ can be given without an ambiguity as shown in the lower-left corner of the top figure of Fig. \ref{fig:FAandGaugino}. If all the Planck scale related contributions are not important, such as in the GMSB scenario, the gaugino mass contribution dominates. For the KSVZ axion, the gaugino contribution to axino mass is shown in the bottom figure of Fig. \ref{fig:FAandGaugino}. Anomaly mediation can contribute too near the Planck scale, but it is subdominant to the gravity mediation contribution.
So axino mass is parametrized as
\dis{
m_{\tilde a}\, &=   \left(\xi_{\rm goldstino}+\sum_{I= {\rm terms~in~}W} \xi^{\rm anom}_{I}\right)m_{3/2}\\
&\quad + \sum_{a= {\rm gaugino}} \xi_{a\,} m_{1/2\,,a } .\label{eq:axinomall}
}

%%%%%%%%%%%%%%%%%%%%%%%%%%%%%%%%%%%%%%%%%%%%%%%%%%%%%%%%%%%%%%%%%%%%%%%%%%%%%%%%%%%%
\section{Axino mass}

Definition of Goldstino $\tilde{Z}$ can be given without an ambiguity  as shown in the upper-right corner in the top figure of Fig. \ref{fig:FAandGaugino}. But, axino $\tilde{a}$ is defined such that it belongs to a subset of $\tilde{a}\perp \tilde{Z}$ if there are many coaxinos. Therefore, there is no reason that $\tilde{a}$ is the $\vartheta^1$ component of $a$.

In \cite{ChunKN92}, the possiblility of keV axino was discussed in case the superpotential has an {\it accidental symmetry}. The keV \cite{axcosmologykeV} and even eV \cite{axcosmologyeV} range axino masses are possible with some accidental symmetries. The accidental symmetries may forbid the leading order masses of the scales $m_{3/2}$ and $m_{1/2\,,a }$.

In the gravity mediation scenario, $m_{3/2}$ is a TeV scale and the axino mass depends on the K\"ahler potential. With  the axino-Goldstino mixing, it has been calculated recently in \cite{KSeoAxinomass12}.

In the gauge mediation scenario, $m_{3/2}$ is negligible and the axino-Goldstino mixing does not give a  significant contribution. Then, the loops may give the dominant contribution. But the accidental symmetry may forbid diagrams of the form in the bottom figure of Fig. \ref{fig:FAandGaugino}. The superpotential may introduce a nonrenormalizable term suppressed by $M_P$, and then the expansion parameter is $f_a/M_P\sim 10^{-7}$. Thus, the axino mass diagram of Fig.  \ref{fig:FAandGaugino} is further suppressed by $\sim 10^{-7}$ and we expect $10\,\gev\cdot 10^{-7}\simeq 1 ~$keV. If it is further suppressed, then the estimated axino mass is of order $10^{-3}\,$eV.

%%%%%%%%%%%%%%%%%%%%%%%%%%%%%%%%%%%%%%%%%%%%%%%%%%%%%%%%%%%%%%%%%%%%%
\subsection{Gaugino contribution to the KSVZ axino mass}\label{subsec:KSVZ}
%%%%%%%%%%%%%%%%%%%%%%%%%%%%%%%%%%%%%%%%%%%%%%%%%%%%

In the KSVZ approach, one introduces the heavy quark fields $Q_L$ and $Q_R$ in the superpotential as  \cite{KSVZ79},
\begin{eqnarray}
W_{\rm KSVZ}=m^3\Theta\, e^{A/f_a} &+{f_Q} Q_L \overline{Q}_R \,
       \varphi  \, e^{A/f_a }  . \label{Wksvz}
\end{eqnarray}
The PQ symmetry is given near the $\epsilon$ point in the bottom figure of Fig. \ref{fig:AxionShift}, with $\Gamma(\overline{Q}_L)=-1/2, \Gamma( {Q}_R)=-1/2,$ and $\Gamma(X)=1 $. Near $\epsilon$, there is no $\varphi$ type field. But near $N_{\rm DW}$, $\overline{Q}_L$ and $Q_R$ are not of the $\varphi$ type,  only $X$ is a $\varphi$ type field, and $Q$ obtains the heavy quark mass $m_Q=f_Q \langle\varphi(X)\rangle$.

It can be rephrased as follows. After integrating out heavy scalars by $\varphi=f_a$, for the heavy quark interaction with $A$  we have  $m_QQ_L\overline{Q}_Re^{ A/f_a}$. Technically, we loose the PQ quantum number information of heavy quarks since they do not have a $\varphi$ type component but only the phase dependence by the original PQ charges. These phases can be rotated away by redefining the phases of $Q_L$ and $Q_R$. This heavy quark interaction with $A$ generates the two loop mass of order 10 GeV as shown in the bottom figure of Fig. \ref{fig:FAandGaugino}.

%%%%%%%%%%%%%%%%%%%%%%%%%%%%%%%%%%%%%%%%%%%%%%%%%%%%%%%%%%%%%%%%%%%%%%%
\begin{table}
\begin{center}
\begin{tabular}{|c|c|c|c|c|c|c|c|c|c| }
\hline
Model    & $S_1$ & $S_2$ & $Q_L$ & $\overline{Q}_R$
&$H_d$ &$H_u$& $q_L$& $D^c_R$ & $U^c_R$ \\
\hline
{\rm KSVZ}  & $1$&$-1$&$-\frac12 $&$-\frac12 $& 0& 0& 0&0&0\\
{\rm DFSZ}  & $1$&$-1$& 0 & 0&$-1 $ & $- 1$& $\ell$& $1-\ell$ &$1-\ell$\\
\hline
\end{tabular}
\caption{The PQ charge assignment $Q$. $Q_L$ and $\overline{Q}_R$ denote
new heavy  quark multiplets.   }
\label{table:PQcharge}
\end{center}
\end{table}
%%%%%%%%%%%%%%%%%%%%%%%%%%%%%%%%%%%%%%%%%%%%%%%%%%%%%%%%%%%%%%%%%%%%

%%%%%%%%%%%%%%%%%%%%%%%%%%%%%%%%%%%%%%%%%%%%%%%%%%%%%%%%%%%%%%%%%%%%%
\subsection{Gaugino contribution to the DFSZ axino mass}\label{subsec:DFSZ}
%%%%%%%%%%%%%%%%%%%%%%%%%%%%%%%%%%%%%%%%%%%%%%%%%%%%

In the DFSZ framework, the SU(2)$_L\times $U(1)$_Y$ Higgs doublets carry PQ charges
and thus the light quarks are also charged under U(1)$_{\rm PQ}$ \cite{DFSZ81}.
The charge assignment is shown in Table~\ref{table:PQcharge}.
So, the superpotential is written as

\begin{eqnarray}
 W_{\rm DFSZ}=W_{\rm PQ}+ \frac{f_s}{M_P} S_1^2 H_d H_u,
\end{eqnarray}
where  $H_d H_u\equiv \epsilon_{\alpha\beta}H_d^\alpha H_u^\beta$. Integrating out $S_1$, we have
\dis{
 W_{\rm DFSZ}&=  \mu  e^{2A/f_a}\varphi(H_d) \varphi(H_u)e^{-2A/f_a}\\
 &\quad + f_u q_L e^{ {\ell}  \theta} u_R^c e^{(1-{\ell})\theta} \varphi(H_u)e^{-  A/f_a} \\
  &\quad +f_d q_L e^{ {\ell} \theta} d_R^c e^{(1- {\ell} )\theta} \varphi(H_d)e^{-  A/f_a}   .\label{eq:DFZSsuper}
}
For the quarks, they do not contain the $\varphi$ type fields since they do not contribute to $V_a^2$ of Eq. (\ref{eq:axiondefSUSY}) and their phase is just a phase parameter $\theta$. This $\theta$ can be removed by
redefining the phases of quarks, and we obtain
\begin{eqnarray}
 W_{\rm DFSZ}= \mu   \frac{v_uv_d}{2}
 +( m_t t_L  t_R^c   + m_b b_L  b_R^c  +\cdots) e^{ A/f_a}  \label{eq:DFZSsuperpotent}
\end{eqnarray}

%%%%%%%%%%%%%%%%%%%%%%%%%%%%%%%%%%%%%%%%%%%%%%%%%%%%%%%%%%%%%%%%%%%%%
\section{Gravity mediation, axino-Goldstino mixing, and axino mass}

In the Higgs mechanism, after the gauge symmetry is broken, there appears the exactly massless longitudinal component of the gauge boson. There exists the massless pseudoscalar direction in the mass matrix of pseudoscalar fields. In the super-Higgs mechanism, correspondingly there appears the exactly massless spin-$\frac12$ direction, the {\it Goldstino} direction, which is absorbed to the spin-$\frac32$ gravitino to render it mass $m_{3/2}$. So, the mass matrix for the spin-$\frac12$ chiral fields has the $m=0$ direction which is interpreted as the Goldstino direction.

The PQ symmetry must be respected in $W$ and $K$. In $W$, we assumed that the PQ symmetry is linearly realized. In the K\"ahler potential, complex scalar fields $\phi_i$ and their complex conjugates $\phi_i^*$ appear. The axion superfields $A$ appear in the exponent. Therefore, the exponent must not involve $a$ explicitly, \ie $K$ contains only the $A$ function of the form $A+A^*$.

If the gravity mediation dominates, then the axino mass is of order $m_{3/2}$. Without the axino-Goldstino mixing in the K\"ahler potential, the axino direction is the same as that of axion and the superpotential determines the axino mass. It means that axino mass arises from loop diagrams as in Fig. \ref{fig:FAandGaugino}. Therefore, without the axino-Goldstino mixing in the K\"ahler potential, axino mass is not going to be larger than 10 GeV. Thus, a very heavy axino mass is possible only if there is a significant $A-Z$ mixing in the K\"ahler potential.

Chun and Lukas studied axino with the minimal K\"ahler form  \cite{ChunLukas95}. Here we go beyond the minimal K\"ahler form, work with the PQ symmetry realized in the Nambu-Goldstone manner, and include the effects of F-terms of the PQ charged fields which affect the axino component.

The lowest order terms in the K\"ahler potential with some mixing with SUSY breaking coaxino $C$ are
\dis{
K= \frac12 &(A+A^*)^2 +\epsilon (A+A^*)(C+C^*)\\
&  +CC^*+M(A+A^*).\label{eq:MixKaehler}
}
The SUSY breaking is parametrized by an auxilliary holomorphic constant $\Theta$,
$\Theta=1+ m_{\rm S}\vartheta^2$. If there are coaxions then the superpotential can be taken as $W(C)=\frac{C^4}{M_P}\Theta+\cdots$, with $\langle W(C)\rangle=M^3\sim (10^{13}\,\gev)^3$.

The simplest case for axino-Goldstino mixing is for one co-axino case, just the Goldstino. Then we consider a $2\times 2$ mass matrix of the chiral spinor fields.
Here, we require three plausible conditions:
\begin{itemize}
\item[({\it i})] The vanisihing CC condition,
\begin{eqnarray}
\hskip -1cm G^{i{\bar j}}G_iG_{\bar j}=3M_P^2,\label{eq:Cond1}
\end{eqnarray}
where where $G=K+M_P^2 \ln|W|^2$.
\item[({\it ii})]  The vacuum stabilization condition,
\begin{eqnarray}
\hskip -1cm G^{j {\bar k}}G_{\bar k}\nabla_i G_j+G_i=0.\label{eq:Cond2}
\end{eqnarray}

\item[({\it iii})]  For the U(1) invariance condition, we use
\begin{eqnarray}
&K&=K(A+A^*, C,C^*)  \\
 &W&=\Theta \, e^{\alpha A/f_a}W(C).\label{eq:WofA}
\end{eqnarray}
If there are more than one coaxino, we have
$W=W(C)\, e^{\alpha A/f_a}\times  e^{\alpha A_1/f_1}\times\cdots$.
The superpotential in (\ref{eq:WofA}) preserves the shift symmetry of $A$ since
in  $G=K+\ln|W|^2$, the $|W|^2$ part is read as
$|W|^2=|W(C)|^2\, \Theta\, e^{\alpha(A+A^*)/f_a}.$

\end{itemize}
We considered the axino mass matrix given by
\begin{eqnarray}
     m=m_{3/2}\left[\nabla_iG_j+\frac13G_iG_j\right]
\end{eqnarray}
for two classes of $\langle C\rangle=0$ and $\langle C\rangle\ne 0$ \cite{KSeoAxinomass12}.

In Ref. \cite{KSeoAxinomass12}, we studied two cases for $G_A=0$ and $G_A\ne 0$ in some detail and found that there is no clear lower bound on the axino ($\tilde a\perp (m=0~\rm component)$) mass. However, the expression shows the plausible lower limit of $m_{\tilde a}\gtrsim m_{3/2}$. For example, Case for $G_A=0$ and $G^A\ne0$  is studied with
\dis{
K &=\frac12(A+A^*)^2+CC^*+\epsilon(A+A^*)(C+C^*),\\
 &\quad W=e^{\alpha A/f_a}W(C).\label{eq:Kaehlerzero}
}
The reason we can study the case in some detail is that at the quadratic level, the K\"ahler potential is fixed as given in Eq. (\ref{eq:Kaehlerzero}). In this case, $m_{\tilde a}\ge m_{3/2}$. This simple calculation is in the interaction picture.
In addition, kinetic mixing can be taken into account also. Our simple result is that the axino mass is of order the gravitino mass, and probably larger than the gravitino mass. This detail study is for the one coaxino case. Many coaxinos can be different from this result.

%%%%%%%%%%%%%%%%%%%%%%%%%%%%%%%%%%%%%%%%%
\section{A new simple parametrization of the CKM matrix}

The discussion on the strong CP is not separable from the discussion of the weak CP violation \cite{KimRMP10}. Recently, it has been pointed out that a new parametrization of the CKM matrix $V_{\rm CKM}~(\equiv V~\rm below)$ with one row (or column) real is very useful to scrutinize the physical effects of the weak CP violation. Then, the elements of the determinant directly give  the weak CP phase \cite{KimSeo11}. The physical significance of the weak CP violation is given by the  Jarlskog determinant which is  a product of two elements of $V$ and two elements of $V^*$ of the CKM matrix, e.g. of the type $V_{12}V_{23}V_{13}^*V_{22}^*$. This Jarlskog determinant is just twice the area of the Jarlskog triangle. In Ref.  \cite{KimSeo12}, we have shown that one easily obtains the Jarlskog determinant from $V$. For example, the  Jarlskog determinant $J$ is the imaginary part of the product of the skew diagonal elements, $J=|{\rm Im\,}V_{13}V_{22}V_{31}|$. To relate the product of four elements of $V$ and $V^*$ to a product of three elements of $V$ can be proved as follows.

%%%%%%%%%%%%%%%%%%%%%%%%%%%%%%%%%%%%%%%%%%%%%%%%%%%%%%%%%%%%%%%%%%%%%%%%%%%%%%%%%%%%
\begin{figure}[!b]
  \begin{center}
  \begin{tabular}{c}
   \includegraphics[width=0.3\textwidth]{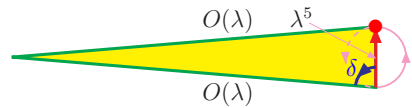}
   \end{tabular}
  \end{center}
 \caption{The Jarlskog triangle. This triangle is for two long sides of $O(\lambda)$. Rotating the $O(\lambda^5)$ side (the red arrow), the CP phase $\delta$ changes.
  }
\label{fig:JTriangle}
\end{figure}
%%%%%%%%%%%%%%%%%%%%%%%%%%%%%%%%%%%%%%%%%%%%%%%%%%%%%%%%%%%%%%%%%%%%%%%%%%%%%%%%%%%%%

If the determinant of $V$ is real, we have ~$ 1=
 V_{11}V_{22}V_{33}-V_{11}V_{23}V_{32} +V_{12}V_{23}V_{31}
 -V_{12}V_{21}V_{33} +V_{13}V_{21}V_{32} -V_{13}V_{22}V_{31}.
$
Multiplying $V_{13}^*V_{22}^*V_{31}^*$ on both sides, we obain
\dis{
V_{13}^* &V_{22}^*V_{31}^*  =|V_{22}|^2V_{11}V_{33}V_{13}^*V_{31}^*-V_{11}
 V_{23}V_{32}V_{13}^*V_{31}^*V_{22}^*\\
 & +|V_{31}|^2V_{12}V_{23}V_{13}^*V_{22}^* -V_{12}V_{21}V_{33}V_{13}^*V_{31}^*
 V_{22}^* \\
 &+|V_{13}|^2V_{21}V_{32}V_{31}^*V_{22}^*-|V_{13}V_{22}V_{31}|^2.\label{eq:detmultV*}
}
We will show that the imaginary part of the left-hand side (LHS) of Eq. (\ref{eq:detmultV*}), \ie  $|{\rm Im}\,V_{31}V_{22}V_{13}|$, is the Jarlskog determinant $J$. Firstly, consider the second term on the right-hand side (RHS), $-V_{11}V_{23}V_{32} V_{13}^*V_{31}^*V_{22}^*$. It contains a factor $V_{32}V_{22}^*$, which is equal to $-V_{31}V_{21}^*-V_{33}V_{23}^*$ by the unitarity of $V$. Then, $-V_{11}V_{23}V_{32}V_{13}^*V_{31}^*V_{22}^*= V_{11}V_{23} V_{13}^*V_{21}^*|V_{31}|^2 +V_{11}V_{33}V_{13}^*V_{31}^*|V_{23}|^2.$ Especially, the second term $V_{11}V_{33}V_{13}^*V_{31}^*|V_{23}|^2$ combines with the first term of Eq. (\ref{eq:detmultV*}), $|V_{22}^2|V_{11}V_{33}V_{13}^*V_{31}^*$, to make $(1-|V_{21}|^2)V_{11}V_{33}V_{13}^*V_{31}^*$. Second, note that the fourth term  on the RHS of Eq. (\ref{eq:detmultV*}), $-V_{12}V_{21}V_{33}V_{13}^*V_{31}^*V_{22}^*$ containing the factor $V_{33}V_{31}^*=-V_{23}V_{21}^*-V_{13}V_{11}^*$. These are used to show \cite{KimSeo12}
\dis{
 V_{13}^* &V_{22}^*V_{31}^*  =(1-|V_{21}|^2)V_{11}V_{33}V_{13}^*V_{31}^*\\
 &\hskip -0.3cm  +V_{11}V_{23} V_{13}^*V_{21}^*|V_{31}|^2 +(1-|V_{11}|^2)V_{12}V_{23}V_{13}^*V_{22}^* \\
 &+|V_{13}|^2(V_{12}V_{21}V_{11}^*V_{22}^*+V_{21}V_{32}V_{31}^*V_{22}^*) \\ &-|V_{13}V_{22}V_{31}|^2.
 \label{eq:detmult3}
}
Let the imaginary part of $V_{11}V_{33}V_{13}^*V_{31}^*$ be $J$. From $V_{11}^*V_{13} +V_{21}^*V_{23}+V_{31}^*V_{33}=0$, we have $|V_{11}|^2|V_{13}|^2$ $+V_{11}V_{23}V_{13}^*V_{21}^*+V_{11}V_{33}V_{13}^*V_{31}^*=0$; so the imaginary part of $V_{11}V_{23}V_{13}^*V_{21}^*$ is $-J$. From $V_{11}V_{31}^*+V_{12}V_{32}^*+V_{13}V_{33}^*=0$, we have $V_{11}V_{33}V_{13}^*V_{31}^*+V_{12}V_{33}V_{32}^*V_{13}^*+|V_{13}^*V_{33}|^2=0$.  And, from $V_{12}^*V_{13}+V_{22}^*V_{23}+V_{32}^*V_{33}=0$,  we have $V_{12}V_{33}V_{32}^*V_{13}^*+V_{12}V_{23}V_{22}^*V_{13}^*+|V_{12}^*V_{13}|^2=0$. These two combine to show that the imaginary part of $V_{12}V_{23}V_{22}^*V_{13}^*$ is $J$. On the other hand, from $V_{11}^*V_{12}+V_{21}^*V_{22}+V_{31}^*V_{32}=0$, we know $V_{21}V_{32}V_{22}^*V_{31}^*+V_{12}V_{21}V_{11}^*V_{22}^*+|V_{21}^*V_{22}|=0$; a similar argument applies to the vanishing
imaginary part of $(V_{21}V_{32}V_{22}^*V_{31}^* +V_{12}V_{21}V_{11}^*V_{22}^*)$. Thus, the imaginary part of the RHS of Eq. (\ref{eq:detmult3}) is
$[(1-|V_{21}|^2)-|V_{31}|^2+(1-|V_{11}|^2)]J=J$.

We can argue that the maximality of the weak CP violation is a physical statement. The physical magnitude of the weak CP violation is given by the area of the Jarlskog triangle. For any Jarlskog triangle, the area is the same. With the $\lambda=\sin\theta_C$ expansion, the area of the Jarlskog triangle is of order $\lambda^6$. In Fig. \ref{fig:JTriangle}, we show the triangle with two long sides of order $\lambda$. Rotating the $O(\lambda^5)$ side (the red arrow of Fig. \ref{fig:JTriangle}), the CP phase $\delta$ and also the area change. The magnitude of the Jalskog determinant is $J\simeq\lambda^6|V_{13}V_{31}/\lambda^6|\sin\delta$. From  Fig. \ref{fig:JTriangle}, we note that the area is maximum for  $\delta\simeq\frac{\pi}{2}$, and the maximality $\delta=\frac\pi{2}$ is a physical statement. The maximal CP violation can be modeled as recently shown in \cite{Kim11}.

%%%%%%%%%%%%%%%%%%%%%%%%%%%%%%%%%%%%%%%%%
\section{Conclusion}

Our result for the axino mass is:
(1) Axino mass can take any value depending on the axion model and SUSY breaking scheme, and (2) We prefer the case for a heavier axino mass compared to the gravitino mass in the gravity mediation.
After properly defining the Goldstino and axion multiplets, we presented our discussion on the axino mass in the most general framework. For only two light superfields of Goldstino and axino, we obtain for $G_A=0$ where $G=K+\ln|W|^2$,
$m_{\tilde a}= m_{3/2}$ with the axino-gravitino mixing parameter  $\epsilon$  in the K\"ahler potential. For $G_A\ne 0$, we showed that the axino mass depends on the details of the K\"ahler potential. But there is another parameter proportional to the gaugino masses, and we can take a wide range of the axino mass for cosmological applications. If the gravity mediation is the dominant one, the axino mass is probably greater than the gravitino mass, but its decay to gravitino is negligible due to the small gravitino coupling. Still, it softens the cosmological gravitino problem \cite{EllisKN84} somewhat as discussed in Ref. \cite{axcosmologyTeV, Huh09}.

\acknowledgments{This work is supported in part by the Korea NRF grant  No. 2005-0093841.}

%%%%%%%%%%%%%%%%%%%%%%%%%%%%%%%%%%%%%%%%%%%%%%%%%%%%%%%%%%%%%%%%%%%%%%%%%%%%%%%%

%%%%%%%%%%%%%%%%%%%%%%%%%%%%%%%%%%%%%%%%%%%%%%%%%%%%%%%%%%%%%%%%%%%%%%%

\def\prp#1#2#3{Phys.\ Rep.\ {\bf #1} (#3) #2}
\def\rmp#1#2#3{Rev. Mod. Phys.\ {\bf #1} (#3) #2}
\def\anrnp#1#2#3{Annu. Rev. Nucl.
Part. Sci.\ {\bf #1} (#3) #2}
\def\npb#1#2#3{Nucl.\ Phys.\ {\bf B#1} (#3) #2}
\def\plb#1#2#3{Phys.\ Lett.\ {\bf B#1} (#3) #2}
\def\prd#1#2#3{Phys.\ Rev.\ {\bf D#1} (#3) #2}
\def\prl#1#2#3{Phys.\ Rev.\ Lett.\ {\bf #1} (#3) #2}
\def\jhep#1#2#3{J. High Energy Phys.\ {\bf #1} (#3) #2}
\def\jcap#1#2#3{J. Cosm. and Astropart. Phys.\ {\bf #1} (#3) #2}
\def\zp#1#2#3{Z.\ Phys.\ {\bf #1} (#3) #2}
\def\epjc#1#2#3{Euro. Phys. J.\ {\bf #1} (#3) #2}
\def\ijmp#1#2#3{Int.\ J.\ Mod.\ Phys.\ {\bf #1} (#3) #2}
\def\mpl#1#2#3{Mod.\ Phys.\ Lett.\ {\bf #1} (#3) #2}
\def\apj#1#2#3{Astrophys.\ J.\ {\bf #1} (#3) #2}
\def\nat#1#2#3{Nature\ {\bf #1} (#3) #2}
\def\sjnp#1#2#3{Sov.\ J.\ Nucl.\ Phys.\ {\bf #1} (#3) #2}
\def\apj#1#2#3{Astrophys.\ J.\ {\bf #1} (#3) #2}
\def\ijmp#1#2#3{Int.\ J.\ Mod.\ Phys.\ {\bf #1} (#3) #2}
\def\apph#1#2#3{Astropart.\ Phys.\ {\bf B#1}, #2 (#3)}
\def\mnras#1#2#3{Mon.\ Not.\ R.\ Astron.\ Soc.\ {\bf #1} (#3) #2}
\def\apjs#1#2#3{Astrophys.\ J.\ Supp.\ {\bf #1} (#3) #2}
\def\aipcp#1#2#3{AIP Conf.\ Proc.\ {\bf #1} (#3) #2}

%%%%%%%%%%%%%%%%%%%%%%%%%%%%%%%%%%%%%%%%%%%%%%%%%%%%%%%%%%%%%%%%%%%%%%%

\end{document}